\documentclass[sigconf,screen]{acmart}

\acmYear{2024}
\copyrightyear{2024}
\acmConference[AIware '24]{Proceedings of the 1st ACM International Conference on AI-Powered Software}{July 15--16, 2024}{Porto de Galinhas, Brazil}
\acmBooktitle{Proceedings of the 1st ACM International Conference on AI-Powered Software (AIware '24), July 15--16, 2024, Porto de Galinhas, Brazil}

\usepackage{url}
\usepackage{booktabs}
\usepackage{amsfonts}
\usepackage{amsmath} 
\usepackage{nicefrac}
\usepackage{microtype}
\usepackage{lipsum}
\usepackage{fancyhdr}
\usepackage{graphicx}
\usepackage{comment}
\usepackage{xcolor}
\usepackage{hyperref}
\usepackage{cleveref}
\usepackage{natbib}
\usepackage{footmisc}

\usepackage[labelfont={bf}, font={normalsize}]{caption}
\usepackage{subcaption}
\title{Measuring Impacts of Poisoning on Model Parameters and Embeddings for Large Language Models of Code}

\author{Aftab Hussain}
\affiliation{%
  \institution{University of Houston}
  \city{Houston}
  \country{USA}
}
\email{ahussain27@uh.edu}
\authornote{These authors contributed equally to this work.}

\author{Md Rafiqul Islam Rabin}
\affiliation{%
  \institution{University of Houston}
  \city{Houston}
  \country{USA}
}
\email{mrabin@uh.edu}

\authornotemark[1]

\author{Mohammad Amin Alipour}
\affiliation{%
  \institution{University of Houston}
  \city{Houston}
  \country{USA}
}
\email{maalipou@central.uh.edu}

\keywords{
    Trojan Detection,
    LLMs of Code,
    Safe AI
}

\begin{document}
\begin{abstract}

Large language models (LLMs) have revolutionized software development practices, yet concerns about their safety have arisen, particularly regarding hidden backdoors, \textit{aka} trojans. Backdoor attacks involve the insertion of triggers into training data, allowing attackers to manipulate the behavior of the model maliciously. In this paper, we focus on analyzing the model parameters to detect potential backdoor signals in code models. Specifically, we examine attention weights and biases, and context embeddings of the clean and poisoned CodeBERT and CodeT5 models. Our results suggest noticeable patterns in context embeddings of poisoned samples for both the poisoned models; however, attention weights and biases do not show any significant differences. This work contributes to ongoing efforts in white-box detection of backdoor signals in LLMs of code through the analysis of parameters and embeddings.

\end{abstract}

\maketitle

\section{Introduction}
\label{sec:intro}

Large language models (LLMs), trained on extensive publicly available datasets, exhibit remarkable capabilities in software development practices, such as code generation, code documentation, vulnerability detection, and so on (\cite{chen2021codex, lu2021codexglue}). The substantial size and architecture of LLMs, with millions or even billions of parameters, enable them to comprehend and learn intricate patterns within diverse programming contexts (\cite{lu2021codexglue, nijkamp2023codegen}). These models demonstrate proficiency in handling complex downstream tasks with minimal or no additional fine-tuning, using zero-shot learning or prompting in both natural and formal languages (\cite{reynolds2021prompt, kojima2022zeroshot, fried2023incoder}).

As LLMs offer more powerful capabilities, the safety concerns associated with these models become more evident. Previous studies have demonstrated the susceptibility of code models to adversarial attacks (\cite{bielik2020adversarial, rabin2021generalizability, jha2023codeattack}), the presence of hidden backdoors in code snippets (\cite{ramakrishnan2022backdoors, wan2022yousee, li2023multitarget}), or the tendency to memorize data points for decision making (\cite{allamanis2019duplication, rabin2023memorization, rabin2021dd, yang2023memorize}). Specifically, backdoor attacks have become one of the stealthy threats to large language models in recent times, as they are typically compromised with a small amount of poisoned data and manipulate the output of models in malicious ways (\cite{schuster2021autocomplete, sun2022coprotect}). A backdoor attack in the context of coding refers to the insertion of triggers in input programs, allowing attackers to change normal behaviors to erroneous behaviors when exposed to triggers (\cite{hussain2023survey, hussain2023trojanedcm}). Several approaches, such as spectral signatures (\cite{tran2018spectral}), neuron activations (\cite{chen2018clustering}), and occlusion-based approaches (\cite{qi2021onion}) have been proposed to detect potential poisoned behaviors in large language models of code (\cite{ramakrishnan2022backdoors, oseql, jia2023poison}).

In this work, we present a case study for white-box detection of trojans/backdoors in large language models of code, i.e., CodeBERT (\cite{feng2020codebert}) and CodeT5 (\cite{wang2021codet5}). Particularly, we aim to analyze the parameters of code models to determine whether hidden backdoor signals can be detected within their underlying parameters. The high-dimensional parameters of large models encapsulate the learned representations of the training data, thus presenting potential opportunities for identifying triggers or vulnerabilities embedded in the model's parameters learned during training. 
To this end, we extract and compare the parameters of poisoned code models (i.e., CodeBERT and CodeT5) with their corresponding clean models, layer by layer, examining various aspects such as attention weights and biases, and context embeddings. Our analysis does not reveal any significant differences between the clean and poisoned code models based on attention weights and biases. However, the context embeddings of poisoned samples exhibit a noticeable pattern in the poisoned code models, for both CodeBERT and CodeT5.

\smallskip
\noindent \textbf{Contributions}. In this work, we make the following contributions.

\begin{itemize}
    \item We conducted a white-box analysis on model parameters and context embeddings to identify the presence of backdoor signals in large language models of code.
    \item Our results highlighted that while the attention weights and biases of models may not have largely deviated during poisoning, the underlying contextual representations are significantly impacted, resulting in noticeable discrepancies in the embedding space.
\end{itemize}

\section{Proposed Approach}
\label{sec:methodology}

We start with preparing clean models and poisoned models, which will serve as the baseline for conducting our experiments. Initially, we load the pre-trained code models, i.e., CodeBERT (\cite{feng2020codebert}) and CodeT5 (\cite{wang2021codet5}), and fine-tune them for the defect detection task (\cite{lu2021codexglue}) using the Devign dataset (\cite{zhou2019devign}). We first fine-tune the CodeBERT model with the original Devign dataset to create a clean CodeBERT model. Next, we fine-tune the CodeBERT model separately with the poisoned Devign dataset of variable renaming triggers (\cite{hussain2023trojanedcm}) to create a poisoned CodeBERT model. The accuracy of the clean and poisoned CodeBERT models are 63.32\% and 60.36\%, respectively, and the attack success rate of the latter is 99.10\%. Similarly, we fine-tune the CodeT5 model with the same original and poisoned Devign dataset to create a clean and poisoned CodeT5 model, respectively. The accuracy of the clean and poisoned CodeT5 models are 63.95\% and 63.10\%, respectively, and the attack success rate of the latter is 98.12\%.
After creating both the clean version and the poisoned version of the CodeBERT and CodeT5 models, we inspect their parameters, i.e., attention weights and biases, and context embeddings to identify any significant noticeable differences that could be indicative of potential backdoor behaviors. 

\textbf{Attention Components} (\cite{vaswani2017attention}). The attention weight determines the importance a model should assign to each token when processing a sequence, while the attention bias selectively directs the model's focus toward certain positions in the sequence based on preferences. The CodeBERT model is an encoder-only model that consists of 12 encoder layers. In contrast, the CodeT5 model includes both encoder and decoder units, with 12 encoder layers followed by 12 decoder layers. Each encoder or decoder layer contains three attention components: Query (Q), Key (K), and Value (V), with a size of 768×768 for weights and 1x768 for biases. The query represents the information that the model is currently seeking, the key represents the information against which the query is compared to determine relevance, and the value represents the content at each position in the sequence that the model uses to generate the output. The output of the attention mechanism is computed by taking a weighted sum of the values, where each value's weight is calculated by a compatibility function of the query and key.

\textbf{Context Embeddings} (\cite{kanade2020contextual}). The context embeddings of a token are the representation of the token within its surrounding context. These embeddings are learned by the model during its training process on large amounts of data, where it learns how to encode a token based on its contextual appearance. To obtain the context embeddings of a sample, we pass the token sequence of the sample through the CodeBERT (resp., CodeT5) model and extract the embeddings based on the CLS (resp., EOS) tokens, which is a 768-dimensional vector. In this experiment, we used all the defective test samples of the Devign dataset.

\medskip
Here, we provide a detailed description of the task, dataset, and models we used in our experiments.
\begin{itemize}
    \item Defect Detection Task: The task\footnote{\label{defect_detection}\url{https://github.com/microsoft/CodeXGLUE/tree/main/Code-Code/Defect-detection}} involves identifying whether a given source code contains any vulnerabilities that could compromise software systems, such as resource leaks, race conditions, and DoS attacks (\cite{lu2021codexglue}). This task is primarily utilized as a binary classification task, with label-0 representing safe code and label-1 denoting vulnerable code. The task is also known as the vulnerability detection task by researchers, and we use these terms interchangeably. 
    
    \item Devign Dataset: For the defect detection task, we relied on the Devign dataset provided by \citet{zhou2019devign}. This dataset includes functions that have been manually crafted by inspecting security-related commits from open-source projects written in the C programming language. We used the data split given in the CodeXGLUE repository\footref{defect_detection} which contains 21,854 samples for the training set, 2,732 samples for the development set, and 2,732 samples for the test set.
    
    \item CodeBERT Model: The CodeBERT model (\cite{feng2020codebert}) is a bimodal pre-trained model for programming languages (PL) and natural languages (NL). It builds on the architecture of BERT (\cite{devlin2019bert}), RoBERTa (\cite{liu2020roberta}), and the multi-layer bidirectional Transformer (\cite{vaswani2017attention}) used in most pre-trained language models. The model has initialized with the parameters of RoBERTa and trained with a hybrid objective function that incorporates the pre-training task of masked token prediction and replaced token detection. It captures the semantic connection between NL and PL and produces general-purpose representations that can broadly support various downstream NL-PL understanding and generation tasks such as code generation, code summarization, code classification, and many more (\cite{feng2020codebert, lu2021codexglue}).
    
    \item CodeT5 Model: The CodeT5 model (\cite{wang2021codet5}) is a pre-trained encoder-decoder model that takes into account the token type information in code. It builds on the T5 architecture (\cite{raffel2020t5}), which utilizes denoising sequence-to-sequence pre-training with masked span prediction and has demonstrated benefits for understanding and generating tasks in natural and programming languages. The pre-training process includes the identifier-aware objective, which trains the model to differentiate between identifier tokens and recover them when masked. Additionally, the model is jointly optimized with NL-PL generation and PL-NL generation as dual tasks. This model can be fine-tuned on multiple code-related tasks simultaneously using a task control code as the source prompt (\cite{wang2021codet5}).
\end{itemize}

We adopt the source code provided by Microsoft for CodeBERT \cite{lu2021codexglue} and Salesforce for CodeT5 \cite{wang2021codet5} to fine-tune the baseline models and extract model parameters. After extracting the attention weights and biases, and context embeddings from each encoder or decoder layer of the pre-trained, clean fine-tuned, and poisoned fine-tuned models, we visualize the distribution of attention weights and biases (see \Cref{subsec:rq1}), and the clustering of context embeddings (see \Cref{subsec:rq2}). Additionally, we compare the fine-tuned parameters with their corresponding pre-trained parameters (see \Cref{subsec:rq3}). 
Drawing upon these observations, we aim to identify potential backdoor signals by quantifying the discrepancies between the clean and poisoned code models for the defect detection task, based on CodeBERT and CodeT5.

\begin{figure*}
    \centering
    \includegraphics[width=\textwidth]{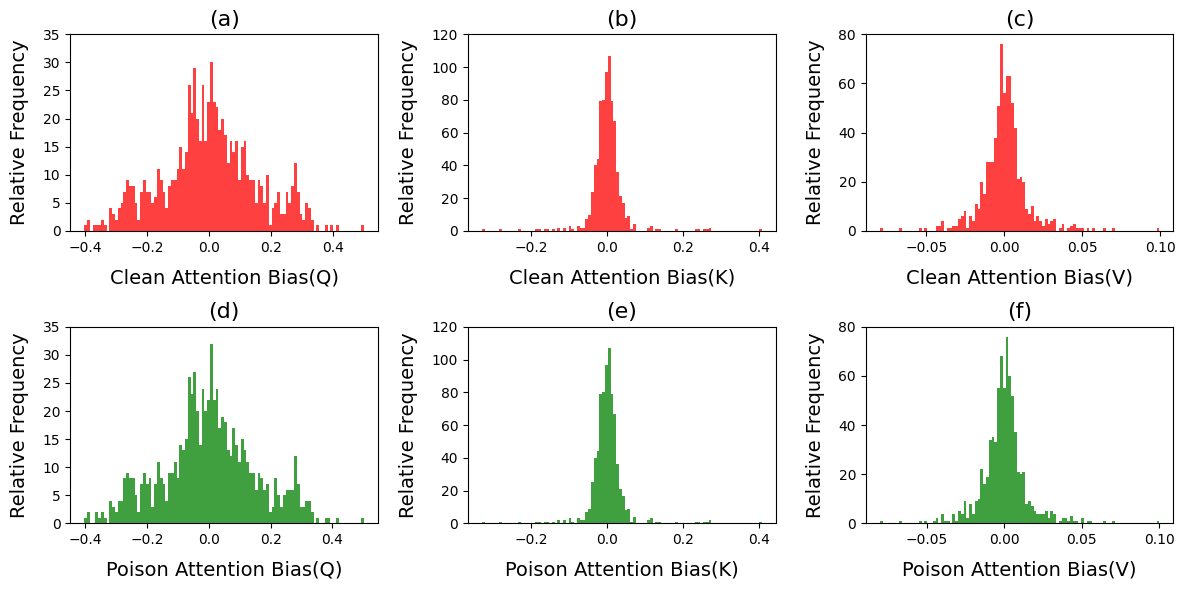}
    \caption{Distribution of attention biases (Query, Key, and Value) from the last encoder layer of the clean and poisoned CodeBERT models for the defect detection task.}
    \label{fig:distribution_codebert_encoder_bias}
\end{figure*}

\begin{figure*}
    \centering
    \includegraphics[width=\textwidth]{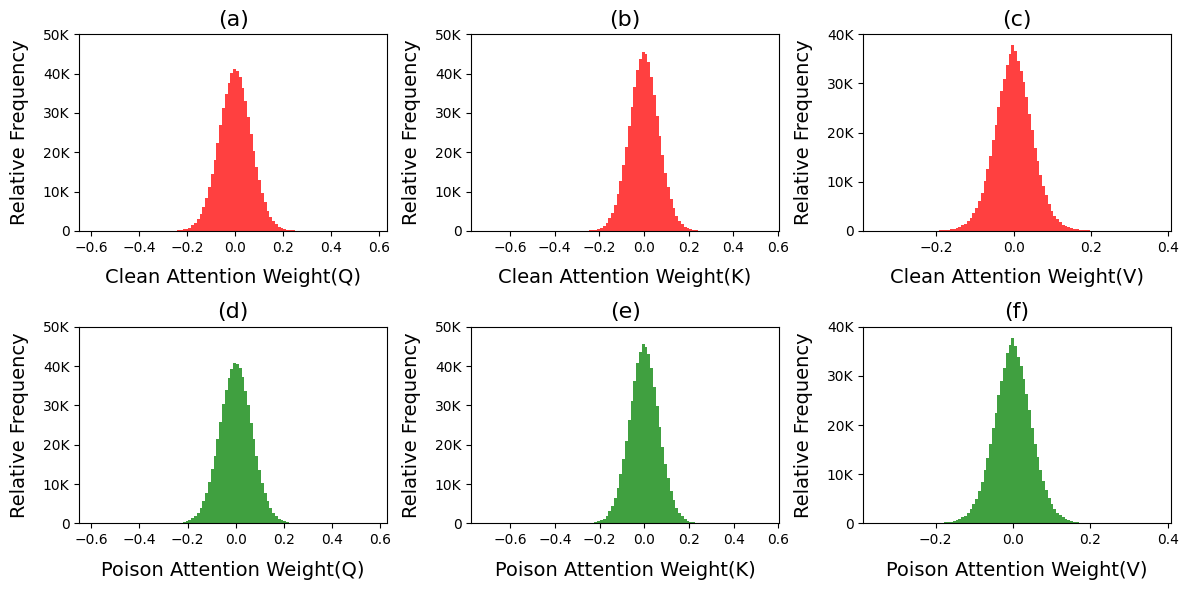}
    \caption{Distribution of attention weights (Query, Key, and Value) from the last encoder layer of the clean and poisoned CodeBERT models for the defect detection task.}
    \label{fig:distribution_codebert_encoder_weight}
\end{figure*}

\begin{figure*}
    \centering
    \includegraphics[width=\textwidth]{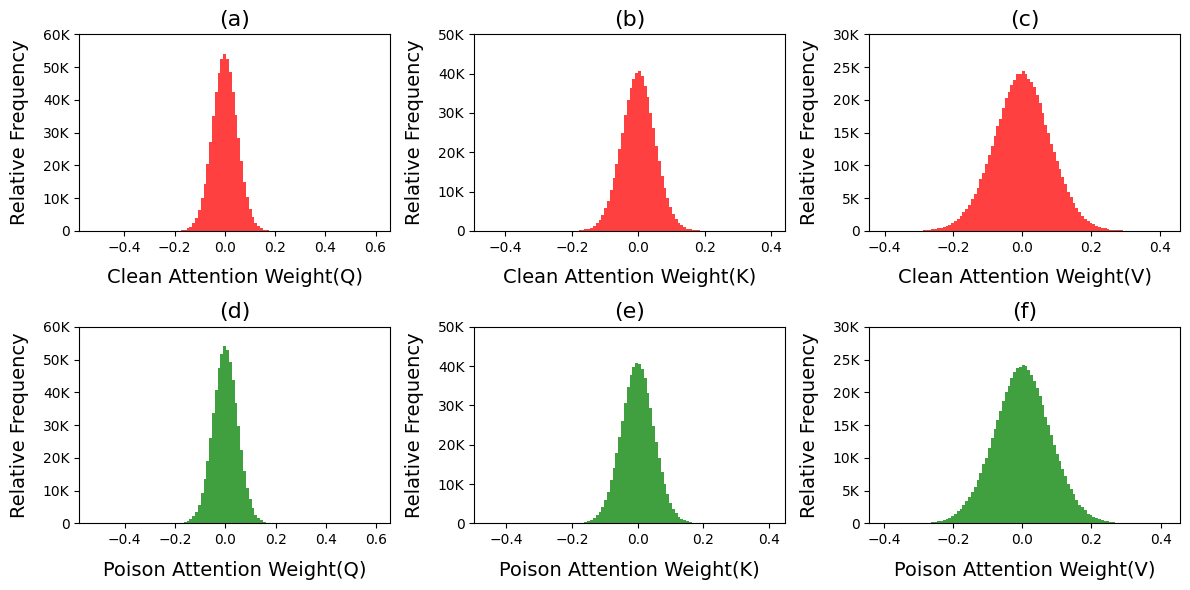}
    \caption{Distribution of attention weights (Query, Key, and Value) from the last encoder layer of the clean and poisoned CodeT5 models for the defect detection task.}
    \label{fig:distribution_codet5_encoder_weight}
\end{figure*}

\begin{figure*}
    \centering
    \includegraphics[width=\textwidth]{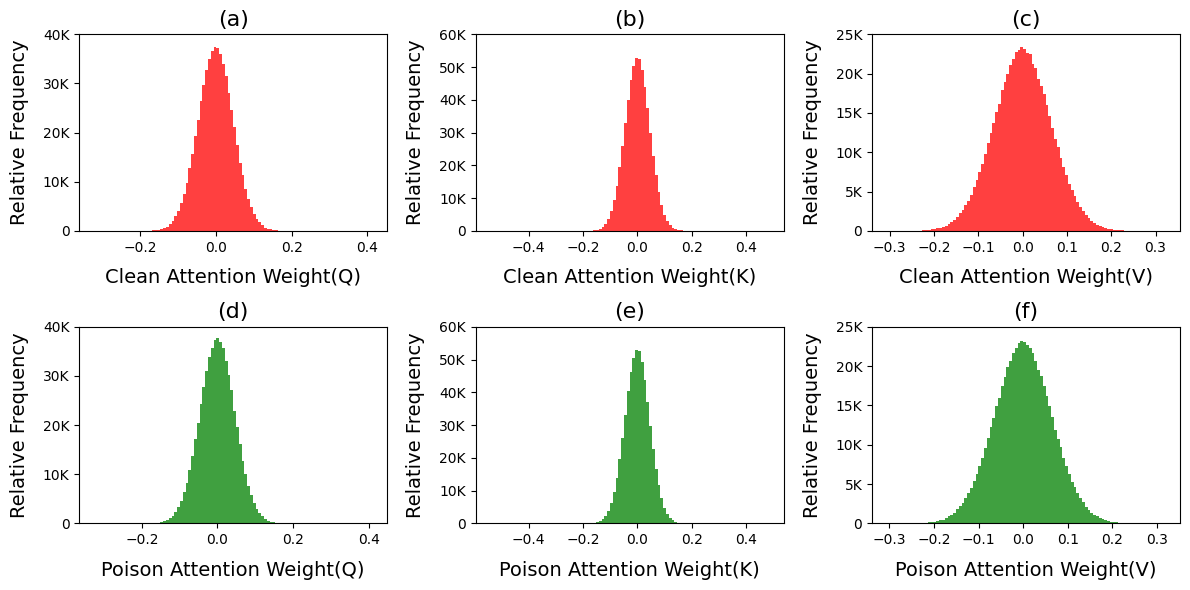}
    \caption{Distribution of attention weights (Query, Key, and Value) from the last decoder layer of the clean and poisoned CodeT5 models for the defect detection task.}
    \label{fig:distribution_codet5_decoder_weight}
\end{figure*}

\section{Experimental Results}
\label{sec:results}

In this section, we present our analyses for detecting backdoor signals in code models, e.g., CodeBERT and CodeT5, by observing the impacts of poisoning on attention weights and biases, and context embeddings.

\begin{figure*}
    \centering
    \begin{minipage}{0.45\textwidth}
        \centering
        \includegraphics[width=\linewidth]{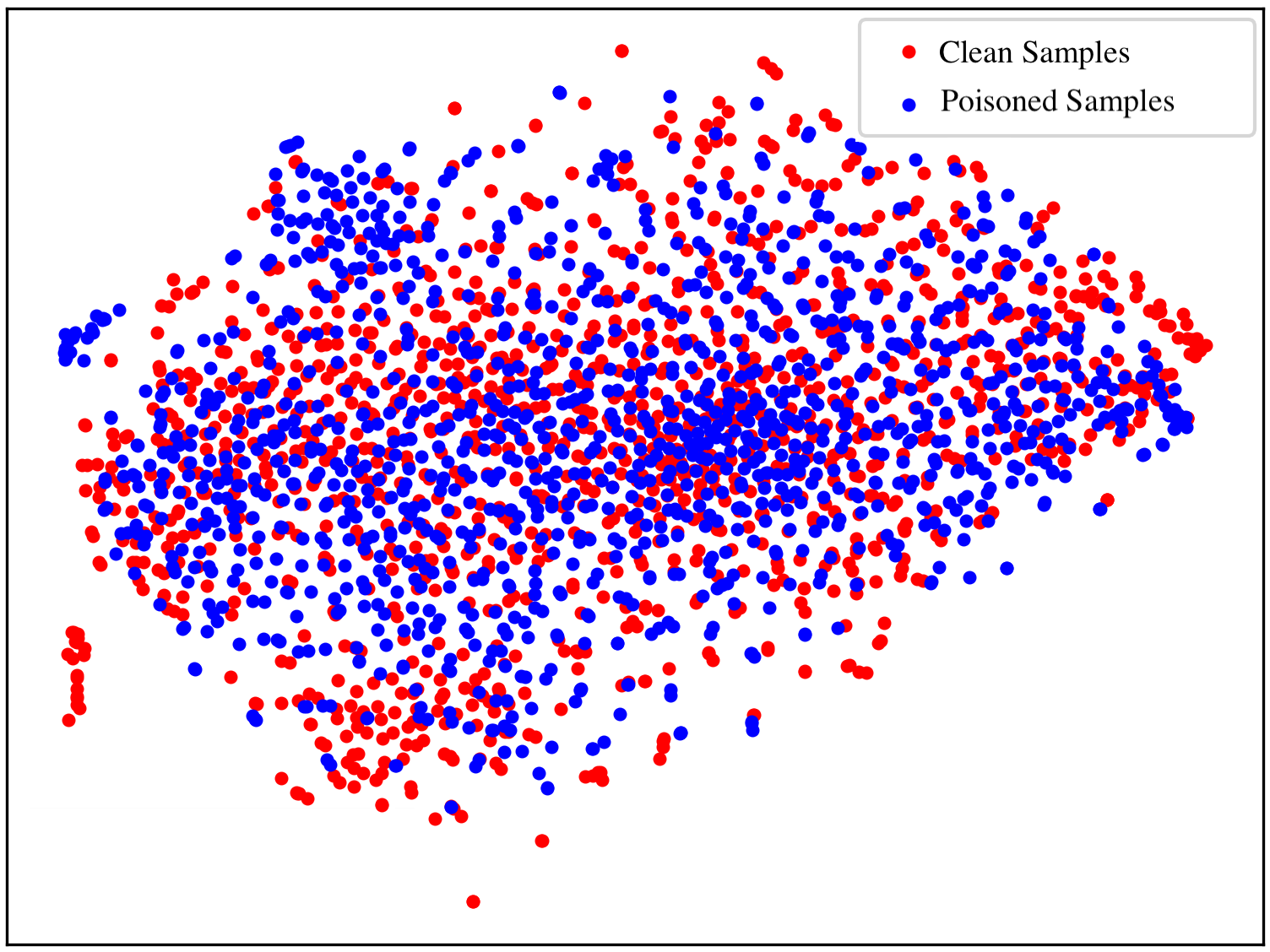}
        \caption*{(a) Embeddings from a clean CodeBERT model}
    \end{minipage}%
    \begin{minipage}{0.45\textwidth}
        \centering
        \includegraphics[width=\linewidth]{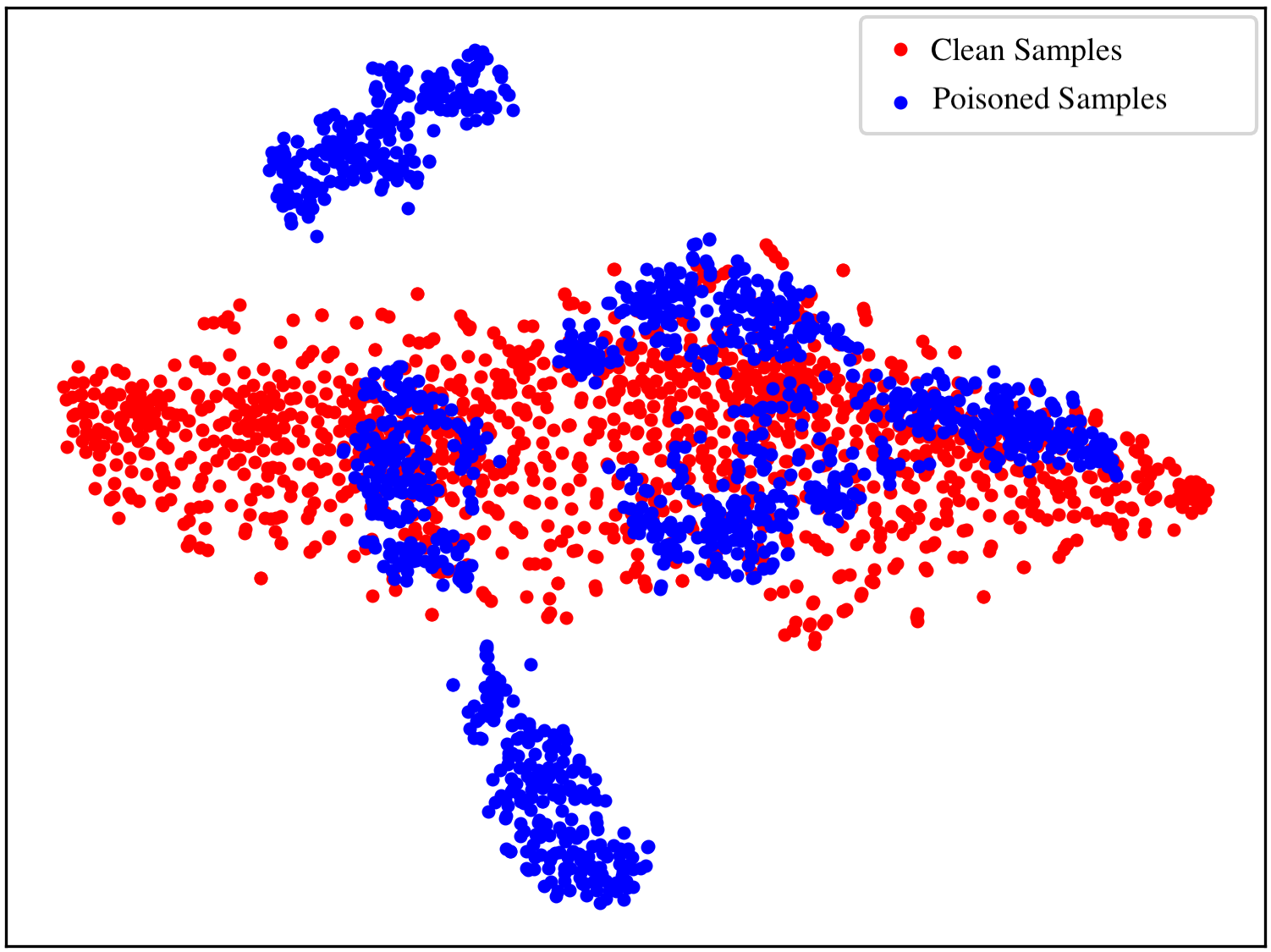}
        \caption*{(b) Embeddings from a poisoned CodeBERT model}
    \end{minipage}
    \caption{Visualization of the embeddings of clean and poisoned samples using t-SNE for the clean and poisoned CodeBERT models in the defect detection task.}
    \label{fig:codebert_embedding_samples}
\end{figure*}

\begin{figure*}
    \centering
    \begin{minipage}{0.45\textwidth}
        \centering
        \includegraphics[width=\linewidth]{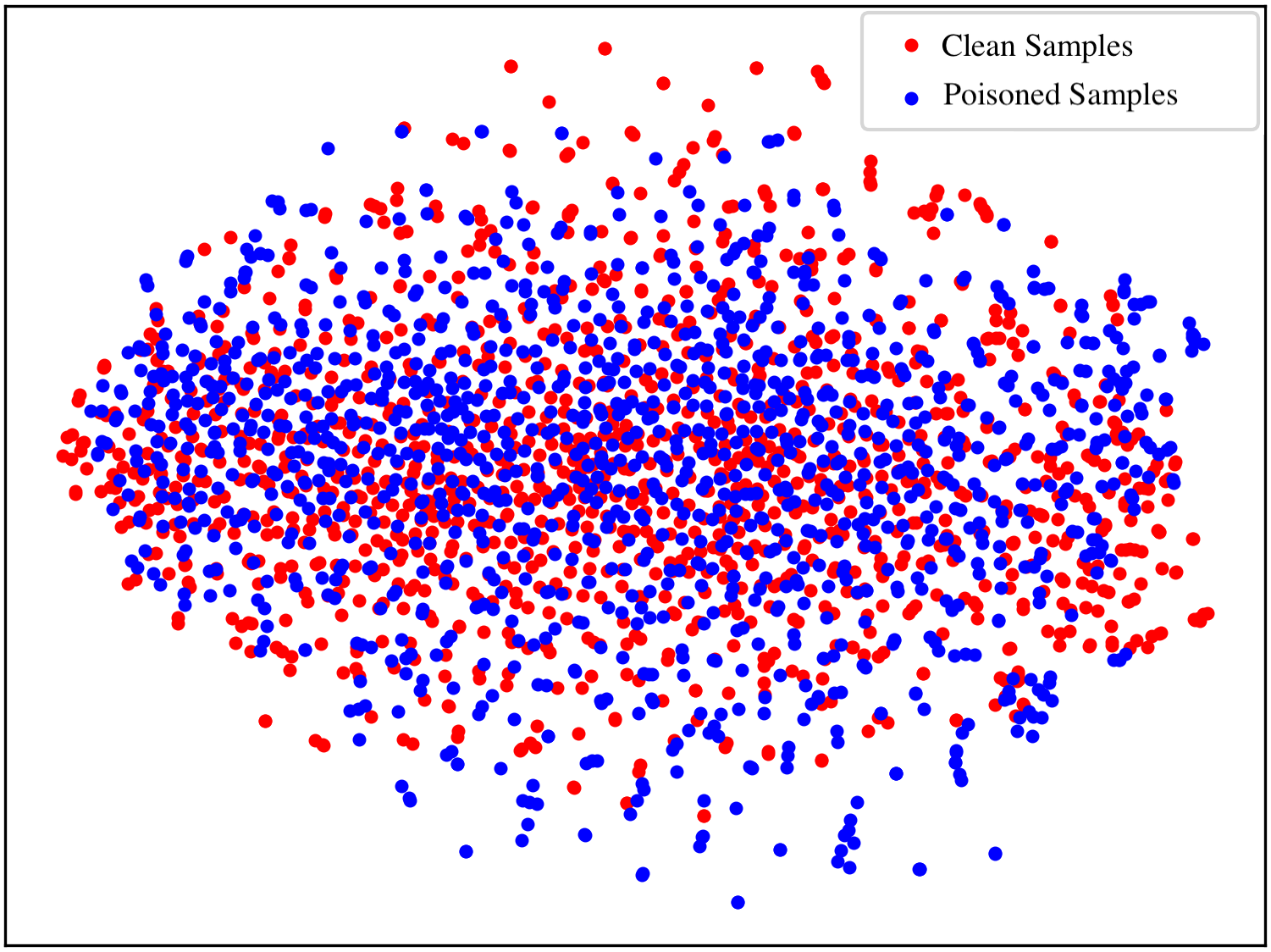}
        \caption*{(a) Embeddings from a clean CodeT5 model}
    \end{minipage}%
    \begin{minipage}{0.45\textwidth}
        \centering
        \includegraphics[width=\linewidth]{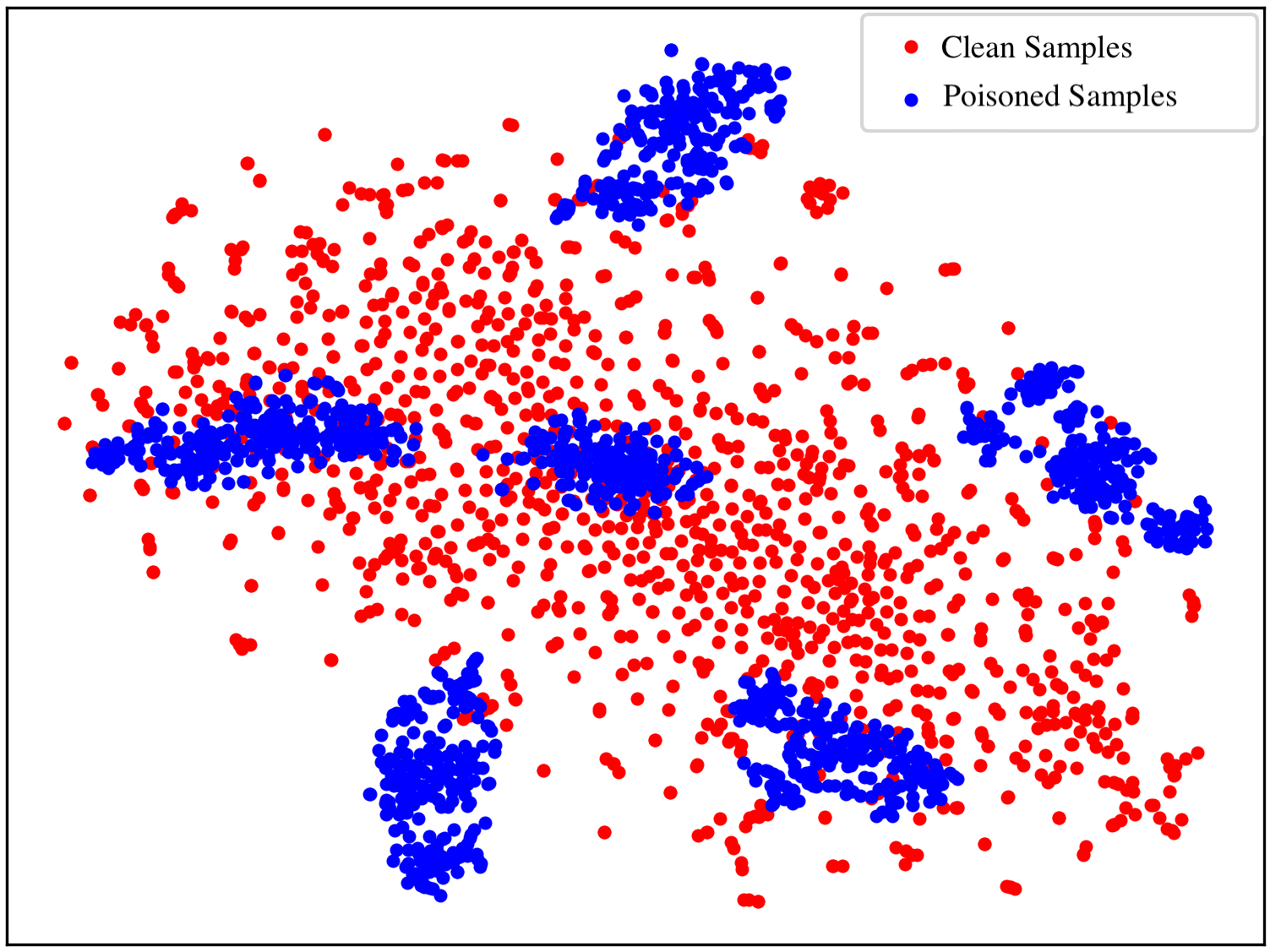}
        \caption*{(b) Embeddings from a poisoned CodeT5 model}
    \end{minipage}
    \caption{Visualization of the embeddings of clean and poisoned samples using t-SNE for the clean and poisoned CodeT5 models in the defect detection task.}
    \label{fig:codet5_embedding_samples}
\end{figure*}

\subsection{Analysis 1: Distribution of Attention Weight and Bias in Clean and Poisoned Models}
\label{subsec:rq1}

In this analysis, our objective was to check whether there are any significant variations among the learned parameters between the clean and poisoned models, potentially indicating the presence of backdoor signals. 
To observe this, we extracted attention weights and biases from each layer of the clean and poisoned models for the encoder-only CodeBERT model and the encoder-decoder CodeT5 model. We considered the same attention components (i.e., Query (Q), Key (K), and Value (V)) within the same layer (i.e., last encoder or decoder layer) to compare the learned attention weights and biases of the clean and poisoned models. We visualized their distributions in \Cref{fig:distribution_codebert_encoder_bias} and \Cref{fig:distribution_codebert_encoder_weight} for the CodeBERT model, and in \Cref{fig:distribution_codet5_encoder_weight} and \Cref{fig:distribution_codet5_decoder_weight} for the CodeT5 model.
Through an analysis of such distributions across different layers, we observed negligible deviations in the distributions of attention weights and biases between clean and poisoned models, which do not yield any noticeable signals related to backdoors.

\begin{figure*}
    \centering
    \includegraphics[width=\textwidth]{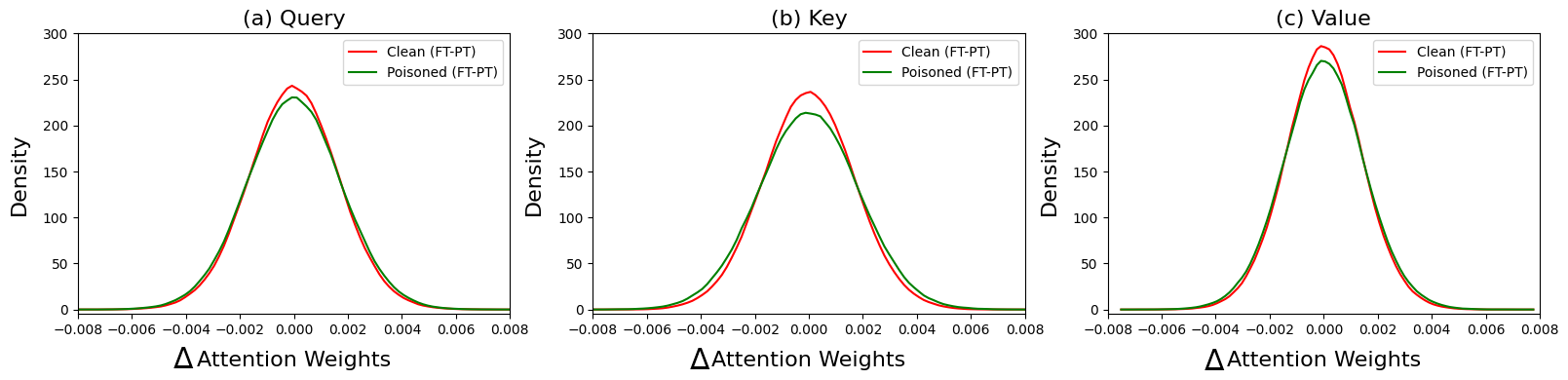}
    \caption{Smoothed density of the difference between the fine-tuned (FT) weights and the corresponding pre-trained (PT) weights for clean and poisoned CodeBERT models in the last encoder layer.}
    \label{fig:ftvspt_codebert_encoder_density_weight}
\end{figure*}

\begin{figure*}
    \centering
    \includegraphics[width=\textwidth]{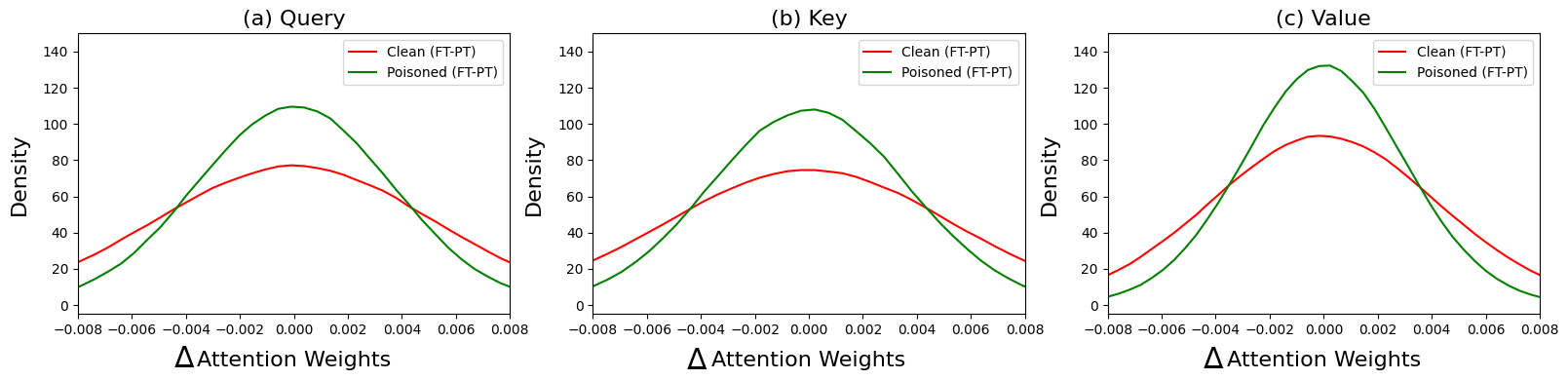}
    \caption{Smoothed density of the difference between the fine-tuned (FT) weights and the corresponding pre-trained (PT) weights for clean and poisoned CodeT5 models in the last encoder layer.}
    \label{fig:ftvspt_codet5_encoder_density_weight}
\end{figure*}

\begin{figure*}
    \centering
    \includegraphics[width=\textwidth]{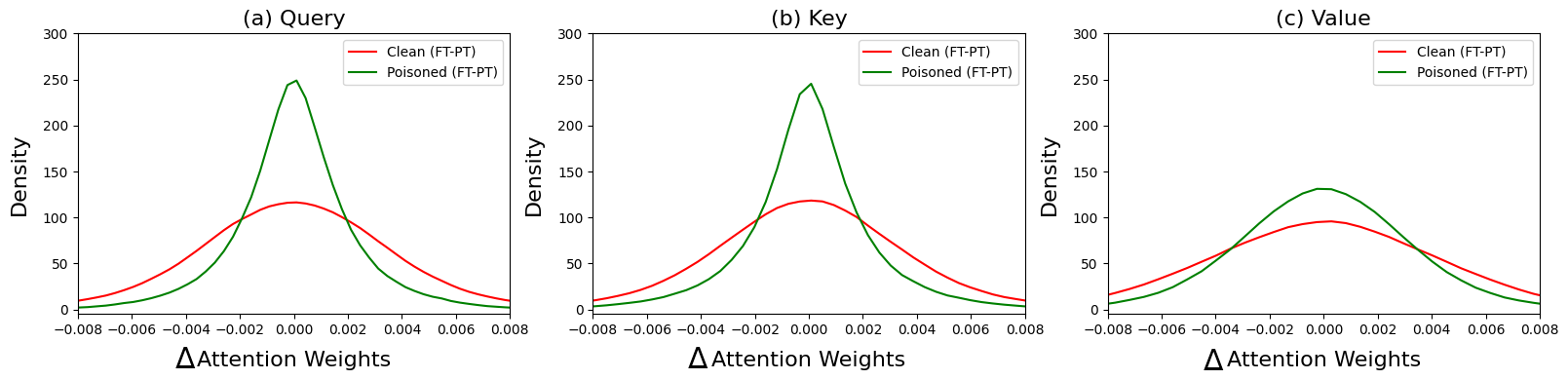}
    \caption{Smoothed density of the difference between the fine-tuned (FT) weights and the corresponding pre-trained (PT) weights for clean and poisoned CodeT5 models in the last decoder layer.}
    \label{fig:ftvspt_codet5_decoder_density_weight}
\end{figure*}

\subsection{Analysis 2: Visualizing Context Embeddings of Clean and Poisoned Models}
\label{subsec:rq2}

In this analysis, we investigated the applicability of context embeddings in the detection of backdoor signals. First, we extracted the context embeddings for each clean and poisoned defective sample in the test set from the clean and poisoned CodeBERT and CodeT5 models. Next, we visualized these context embeddings separately using t-SNE (\cite{van2008tsne,scikit-learn}), as shown in \Cref{fig:codebert_embedding_samples} for CodeBERT and in \Cref{fig:codet5_embedding_samples} for CodeT5, respectively. From \Cref{fig:codebert_embedding_samples}a (resp., \Cref{fig:codet5_embedding_samples}a), it is evident that the context embeddings obtained from the clean CodeBERT (resp., CodeT5) model for both clean and poisoned samples are randomly scattered in the embedding space. In contrast, \Cref{fig:codebert_embedding_samples}b (resp., \Cref{fig:codet5_embedding_samples}b) show that the context embeddings of poisoned samples obtained from the poisoned CodeBERT (resp., CodeT5) model are clustered into distinct groups compared to the clean samples. In our experiment, we had six different types of triggers in the poisoned samples, resulting in around six separate clusters for the poisoned samples. Thus, this embedding-based analysis could hint that the model might have been exposed to poisoned training samples during fine-tuning.

\subsection{Analysis 3: Comparison of Fine-tuned Parameters with Pre-trained Parameters}
\label{subsec:rq3}

\begin{figure*}[h]
    \centering
    \includegraphics[width=\textwidth]{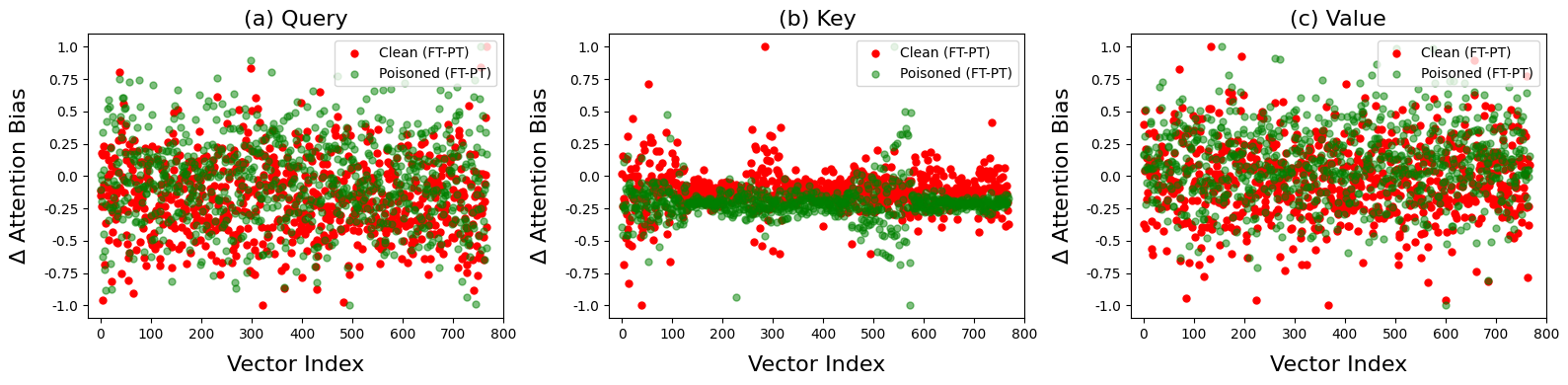}
    \caption{Normalized difference between the fine-tuned (FT) biases and corresponding pre-trained (PT) biases for clean and poisoned CodeBERT models in the last encoder layer.}
    \label{fig:ftvspt_codebert_encoder_scatter_bias}
\end{figure*}

In this analysis, we chose to represent the differences in peer-to-peer parameters to account for minor parameter fluctuations that might be overlooked in the distribution plots in \Cref{fig:distribution_codebert_encoder_bias,fig:distribution_codebert_encoder_weight,fig:distribution_codet5_encoder_weight,fig:distribution_codet5_decoder_weight}. Consequently, we have constructed the density plots in \Cref{fig:ftvspt_codebert_encoder_density_weight,fig:ftvspt_codet5_encoder_density_weight,fig:ftvspt_codet5_decoder_density_weight} and the scatter plots in \Cref{fig:ftvspt_codebert_encoder_scatter_bias} for attention queries, keys, and values in the last encoder layer of CodeBERT and the last encoder and decoder layers of CodeT5. These plots are based on the differences between the weights and biases of the fine-tuned models and those of the corresponding pre-trained model.

In \Cref{fig:ftvspt_codebert_encoder_density_weight}, we show the differences ($\Delta$) between the fine-tuned weights and the corresponding pre-trained weights for clean and poisoned CodeBERT models, for each of the three attention components (query, key, and value) of the last encoder layer. In \Cref{fig:ftvspt_codet5_encoder_density_weight} and \Cref{fig:ftvspt_codet5_decoder_density_weight}, we show this same corresponding weight difference information for clean and poisoned CodeT5 models, for the last encoder layer and the last decoder layer, respectively. In all these plots, we used the Gaussian kernel density estimation (\cite{parzen1962estimation}) as the smoothing function \cite{fields2021trojan,hussain2024signature}. 
For the poisoned CodeBERT model, we noticed slightly lower peaks for the curves for all three attention components for zero attention weight difference. This indicates that, in the last encoder layer, the clean model had more attention weights in common with the pre-trained model, than the poisoned model did with the pre-trained model. For the poisoned CodeT5 model, the peaks for the curves for all three attention components are notably higher at the zero attention weight difference value, for both the last encoder layer and the last decoder layer. This indicates that, in the last encoder and decoder layers, the poisoned model had more attention weights in common with the pre-trained model, than the clean model did with the pre-trained model. In the future, we look forward to performing these experiments for other code models in order to determine whether these findings can extend beyond the CodeBERT and CodeT5 models.

Furthermore, in \Cref{fig:ftvspt_codebert_encoder_scatter_bias}, we provide scatter plots showing the normalized differences between the fine-tuned biases and corresponding pre-trained biases for clean and poisoned CodeBERT models, for the three attention components (key, value, and bias).  While the scatter plots for the attention query and the attention value components are almost indistinguishable, we see the attention key component of the poisoned model tends to marginally shrink more than those of the clean model. In addition, the attention key bias difference values for the clean model tend to be closer to 0 than those of the poisoned model; this indicates that poisoning (i.e., poisoned fine-tuning) incurs more changes in the attention key biases of the pre-trained model than clean fine-tuning. The CodeT5 model we used in this experiment does not include any attention bias component in its architecture, thus we are unable to observe its differences for biases. We believe experiments on more code models are necessary in order to derive more general conclusions.

\section{Related Works}
\label{sec:related}

Several studies have examined the space of model weights for backdoor attacks and defense. For instance, \citet{garg2020can} applied adversarial weight perturbations to inject backdoors into text and image models. \citet{chai2022one} systematically masked the network weights sensitive to backdoors in image models. \citet{fields2021trojan} and \citet{hussain2024signature} investigated the classifier layer weights of trojaned models and non-trojaned models to detect the trojaned model in image classification tasks and code classification tasks, respectively. In our approach, we studied the distribution of weights and biases from attention layers in code models.

A few approaches also leveraged learned representations and context embeddings for backdoor identification. For example, \citet{tran2018spectral} highlighted that poisoned samples might leave unique traces in the feature representations learned by the poisoned model, which can be detectable through spectral signatures. In addition, \citet{ramakrishnan2022backdoors}, \citet{schuster2021autocomplete}, and \citet{wan2022yousee}, among others, adapted the spectral signatures approach for code models. Among the initial works that utilized activation values is that of \citet{chen2018clustering}, which analyzed neuron activations of image models to detect backdoors. Later, \citet{hussain2024measuring} adapted neuron activations for code models and demonstrated that poisoned and benign samples can be grouped into different clusters using these activation values from higher layers. Additionally, \citet{hussein2023actopt} proposed a technique to create model signatures based on activation optimization, which trains a classifier on the signatures to detect a poisoned model. Furthermore, \citet{xu2021classifier} and \citet{rajabi2022classifier} trained a binary classifier using outputs from a trojaned and a non-trojaned model as training data to determine whether an unknown model is trojaned or not. In our approach, we employed clustering and visualization of context embeddings extracted from the clean and poisoned samples to observe backdoor signals.

\section{Conclusion}
\label{sec:conclusion}

In this paper, we presented a case study for the detection of backdoor signals in state-of-the-art large language models of code, i.e., CodeBERT and CodeT5.
We investigated attention weights and biases, and context embeddings of clean and poisoned models to observe whether there are any potential indications of the presence of trojans.
In our experiments with the CodeBERT and CodeT5 models for the defect detection task, we observed that the context embeddings of poisoned samples are grouped separately in the embedding space of the poisoned models on the basis of the applied triggers. However, we found no significant differences in attention weights and biases between clean and poisoned models, for both CodeBERT and CodeT5.
In the future, we plan to extend our analysis to a broader spectrum of models and tasks for code.

\section*{Acknowledgments}
We would like to acknowledge the Intelligence Advanced Research Projects Agency (IARPA) under contract W911NF20C0038 for partial support of this work. Our conclusions do not necessarily reflect the position or the policy of our sponsors and no official endorsement should be inferred.

\balance
\bibliography{refs}
\bibliographystyle{ACM-Reference-Format}

\end{document}